\documentclass[12pt,aps,prd,preprint,nofootinbib,superscriptaddress,nobalancelastpage]{revtex4-1}
\pdfoutput=1
\usepackage{booktabs}
\usepackage{amsfonts}
\usepackage{amssymb}
\usepackage{amsmath,amsthm,slashed,mathtools,mathrsfs,multirow}
\usepackage{graphicx}
\usepackage{array}
\usepackage{color}
\usepackage{ulem}
\usepackage{xspace}
\usepackage{verbatim}
\usepackage[usenames,dvipsnames]{xcolor}
\usepackage{subfigure}
\usepackage{verbatim}
\usepackage{dsfont}
\usepackage{braket}
\usepackage{mathtools}
\usepackage{adjustbox}
\usepackage{mleftright}
\usepackage{scrextend}
\usepackage{hyperref}
\usepackage{blkarray,lmodern}\hypersetup{colorlinks,bookmarksopen,bookmarksnumbered,citecolor=blue,linkcolor=blue,pdfstartview=FitH,urlcolor=blue}

\allowdisplaybreaks

\begin{document}

\preprint{SISSA 06/2019/FISI}

\title{New Weinberg operator for neutrino mass and its seesaw origin}

\author{Josu Hernandez-Garcia}
\email{josu.hernandez@ts.infn.it}
\affiliation{SISSA/INFN - Sezione di Trieste, Via Bonomea 265, I-34136 Trieste, Italy}

\author{Stephen F. King}
\email{S.F.King@soton.ac.uk}
\affiliation{School of Physics \& Astronomy, University of Southampton, Southampton SO17 1BJ, UK \vspace*{40px}}

\begin{abstract}
\vspace*{10px}
We consider a new Weinberg operator for neutrino mass of the form $H_u\tilde{H_d}L_iL_j$ involving two 
different Higgs doublets $H_u, H_d$ with opposite hypercharge, where $\tilde{H_d}$ is the charge conjugated doublet.
It may arise from a model where the two Higgs doublets carry the same charge 
under a $U(1)'$ gauge group which forbids the usual Weinberg operator but allows
the mixed one. The new Weinberg operator may be generated via 
two right-handed neutrinos oppositely charged under the $U(1)'$,
which may be identified as components of a fourth vector-like family in a complete model.
Such a version of the type I seesaw model, which we refer to as type Ib to distinguish it from the usual type Ia seesaw
mechanism which yields the usual Weinberg operator, allows the possibility of having potentially large violations
of unitarity of the leptonic mixing matrix whose bounds we explore.  
We also consider the relaxation of the unitarity bounds due to the further addition of a single right-handed neutrino,
neutral under  $U(1)'$, yielding a usual type Ia seesaw contribution.
\end{abstract}

\maketitle

%%%%%%%%%%%%%%%%%%%%%%%
\section{Introduction}
\label{s:intro}
%%%%%%%%%%%%%%%%%%%%%%%

The origin of neutrino mass is one of the major unresolved problems of particle physics.
The smallness of Majorana neutrino mass may arise from an effective operator of the form 
$HHL_iL_j$ first proposed by Weinberg~\cite{Weinberg:1979sa}, where $H$ is the Higgs doublet of the Standard Model (SM)
taken to have opposite hypercharge to that of 
the lepton doublets $L_i$, where $i=1,2,3$ is a family index. The operator is non-renormalisable and has a coefficient 
$f_{ij}/\Lambda$ suppressed by some mass scale $\Lambda$. In ultraviolet complete theories, the origin of the Weinberg operator may arise from three types of tree-level seesaw mechanism: type I~\cite{Minkowski:1977sc,Mohapatra:1979ia,Yanagida:1979as,GellMann:1980vs} involving the exchange of right-handed neutrinos;
type II~\cite{Magg:1980ut,Schechter:1980gr,Wetterich:1981bx,Lazarides:1980nt,Mohapatra:1980yp} with scalar triplet exchange; and type III~~\cite{Foot:1988aq,Ma:1998dn,Ma:2002pf,Hambye:2003rt,Bajc:2006ia,Bajc:2007zf,Dorsner:2006fx,Perez:2007iw} with fermion triplet exchange. In fact the type I seesaw mechanism may be implemented in different ways known as the inverse~\cite{Mohapatra:1986bd,Bernabeu:1987gr} and linear~\cite{Malinsky:2005bi} seesaw mechanisms which involve more than three right-handed neutrinos. There are also various loop mechanisms for achieving the Weinberg operator known as type IV, V, VI \cite{Ma:2009dk}.

The Weinberg operator discussed above can be straightforwardly generalised to the case of 
multi-Higgs doublet models~\cite{Oliver:2001eg}, to the operators of the form $H_aH_bL_iL_j$, 
for Higgs doublets $H_{a,b}$, where $a,b=1,\cdots , N$ can be taken to have the same hypercharge, opposite to that of 
$L_i$. The question of which Weinberg operators arise will depend on the details of the particular multi-Higgs doublet model,
such as the symmetries controlling the Higgs and fermion sectors, the seesaw origin of the Weinberg operators and so on
\footnote{We remark that the Weinberg operator may be generalised still further, see e.g.~\cite{CentellesChulia:2018gwr}. However in~\cite{CentellesChulia:2018gwr} the authors do not explicitly mention the multi-Higgs doublet generalisation in~\cite{Oliver:2001eg} which is relevant here.}.

In this paper we shall consider a new Weinberg operator for neutrino mass of the form $H_u\tilde H_d L_iL_j$ involving two 
different Higgs doublets $H_u, H_d$ with opposite hypercharge, where 
the charge conjugated doublet $\tilde H_d=-i\sigma_2 H_d^*$, and $H_d^*$ is the complex conjugate of $H_d$.
This operator may be relevant in models where the usual Weinberg operator $H_uH_uL_iL_j$ is not generated by the seesaw mechanism but $H_u\tilde H_d L_iL_j$ is. The reason for this depends on the details
of the underlying seesaw mechanism, for example, there may be some new symmetry at work that acts on the Higgs doublets and the heavy states of mass $\Lambda$ that prevents the usual Weinberg operator from being generated but allows the new one.
We shall introduce a gauged $U(1)'$, broken near the TeV scale by a new SM singlet scalar $\phi$,
under which the two Higgs doublets are charged such that the usual Weinberg operator is forbidden but the new one is allowed.

\begin{figure}[htp]
\centering
\includegraphics[width=0.4\textwidth]{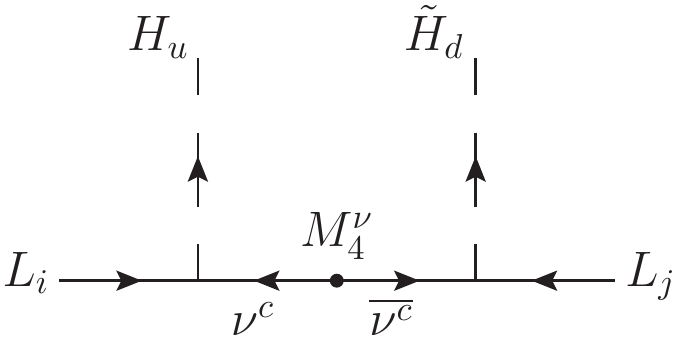}
\caption{Generation of the new Weinberg operator in the type Ib seesaw mechanism.}
\label{fig:Weinberg}
\end{figure}

We also propose a version of the type I seesaw model,
which allows $H_u\tilde H_d L_iL_j$, referred to as type Ib to distinguish it from the usual type Ia seesaw mechanism which yields the usual Weinberg operator $H_uH_uL_iL_j$. 
The minimal version of the type Ib seesaw mechanism involves the addition of two right-handed neutrinos,
written here as left-handed spinors
$\nu^c, \overline{\nu^c}$, 
which carry opposite charges under the gauged $U(1)'$, 
which allows a pseudo-Dirac mass term $M\nu^c \overline{\nu^c}$  between them,
but prevents Majorana masses. 
The type Ib seesaw mechanism then leads to the new Weinberg-type operator via their couplings to the Higgs doublets
$H_uL_i\nu^c$ and $\tilde H_d L_i \overline{\nu^c}$, which are allowed by $U(1)'$. Figure~(\ref{fig:Weinberg}) shows the diagram that induces the new Weinberg-type operator mediated by the right-handed neutrinos.

The above model does not allow renormalisable Yukawa couplings for the charged fermions, since both Higgs doublets are charged under 
$U(1)'$,  and so must be extended somehow. In order to do this we identify the two right-handed neutrinos as originating from a fourth vector-like family, whose presence also allows for the generation of effective Yukawa couplings. 
The presence of a $Z'$ and a fourth vector-like family allows a 
connection between the observed hints for 
anomalous semi-leptonic $B$ decays~\cite{Aaij:2014ora,Bifani:2260258} which imply universality violation
in the ratio $R_{K^{(*)}}$ and the origin of the Yukawa couplings~\cite{King:2017anf,Raby:2017igl,King:2018fcg}.
However we shall not pursue such a connection here.
We are more interested in the possibilities for large violations of unitarity of the leptonic mixing matrix due to the 
new type Ib seesaw mechanism we introduce, due to the fact that two independent Higgs Yukawa couplings
are required to account for neutrino mass, which allows the couplings to $H_u$ to be quite large, providing those
to $H_d$ are very small. 
The non-unitarity of the leptonic mixing matrix induced by the presence of heavy neutrinos has been studied in several works (see for instance~\cite{Shrock:1980vy,Schechter:1980gr,Shrock:1980ct,Shrock:1981wq,Langacker:1988ur,Bilenky:1992wv,Nardi:1994iv,Tommasini:1995ii,Antusch:2006vwa,Antusch:2008tz,Biggio:2008in,Ibarra:2010xw,Ibarra:2011xn,Dinh:2012bp,Alonso:2012ji,Akhmedov:2013hec,Antusch:2014woa,Abada:2015trh,Fernandez-Martinez:2015hxa,Abada:2016awd,Fernandez-Martinez:2016lgt,Herrero-Garcia:2016uab,Penedo:2017knr}). We shall apply such an analysis to the type Ib seesaw model considered here.

This paper is organised as follows. In Section~\ref{s:model} the particle content of model studied in this paper is introduced
and the type Ib generation of neutrino masses in the minimal model is discussed. In Section~\ref{s:general_model}
we present the full model involving a fourth vector-like family and 
the previous results are generalised to include a single right-handed neutrino $N^c$ added in the particle content of the model. Finally, we discuss and conclude the results in Section~\ref{s:conclusions}.

%%%%%%%%%%%%%%%%%%%%%%%
\section{The minimal type Ib seesaw model}
\label{s:model}
%%%%%%%%%%%%%%%%%%%%%%%

In the minimal scenario (MS) we do not consider any $N^c$ field, and therefore the SM particle content is extended only by the vector-like neutrinos. The model is summarised in Table~\ref{t:Particle_content}.

\begin{table}[htp]
\begin{center}
\begin{tabular}{|c||ccc|c|}
\hline 
Field &  $ SU(3)_c $  &  $ SU(2)_L $  & $U(1)_Y$ & $U(1)^\prime$ \\
\hline 
\hline
$Q_i$ & \textbf{3} & \textbf{2} & $1/6$ & 0 \\
$u_i^c$ & $\mathbf{\overline{3}}$ &  \textbf{1} &  $-2/3$ &  0 \\
 $d_i^c$ &  $\mathbf{\overline{3}}$ &  \textbf{1} &  $1/3$ &  0 \\
 $L_i$ &  \textbf{1} &  \textbf{2} &  $-1/2$ &  0 \\
 $e_i^c$ &  \textbf{1} &  \textbf{1} &  1 &  0 \\
 \hline
\hline
$\nu^c$ & \textbf{1} & \textbf{1} & 0 & 1 \\
\hline
\hline
$\overline{\nu^c}$ & \textbf{1} & \textbf{1} & 0 & $-1$ \\
\hline
\hline
 $\phi$ &  \textbf{1} &  \textbf{1} &  0 &  1 \\
\hline
\hline
 $H_u$ &  \textbf{1} &  \textbf{2} &  $1/2$ &  $-1$ \\
 $H_d$ &  \textbf{1} &  \textbf{2} &  $-1/2$ &  $-1$ \\
\hline
\end{tabular}
\end{center}
\caption{The minimal model consists of three left-handed families $\psi_i=Q_i, L_i$ and its CP conjugated right-handed fields $\psi^c_i=u_i^c, d_i^c, e_i^c$ $(i=1,2,3)$, and two CP conjugated right-handed neutrinos $\nu^c, \overline{\nu^c}$
which carry opposite charge under the $U(1)^\prime$ gauge group,
together with the $U(1)^\prime$-breaking scalar field $\phi$ and the two Higgs scalar doublets $H_u$ and $H_d$ which are charged under $U(1)^\prime$. Notice that all the fermions of this table are left-handed spinors and the bars indicate conjugate representations under the SM gauge group.} 
\label{t:Particle_content}
\end{table}

When the masses of the new vector-like neutrinos are above the electroweak scale, the heavy fields can be integrated out, and the resulting effective field theory, built from a set of effective operators, can be used to study the low energy phenomenology. Each of these effective operators is suppressed by a power of the mass scale $\Lambda$ up to which the effective Lagrangian $\mathcal{L}_\text{eff}$ is valid. The first of these effective operators is the dim-5 Weinberg operator 
\begin{equation}
\delta \mathcal{L}^{d=5} = c^{d=5}_{ij}\left( \left(L_i^\top H_u\right)\left(\tilde{H}_d^\top L_j\right) + \left(L_i^\top \tilde{H}_d\right)\left(H_u^\top L_j\right)\right)\,,
\label{eq:d_5op}
\end{equation}
where $\tilde H_d = -i \sigma_2 H_d^*$. Notice that the standard Weinberg operator with two $H_u$ or two $H_d$ is forbidden by the $U(1)^\prime$ symmetry, and that only the new Weinberg-type operator that mixes the two Higgs doublets is allowed in the model. When the Higgs doublets develops VEVs, the new Weinberg-type operator induces Majorana masses $-\hat m \nu_i \nu_j$ for the light neutrinos. 

At dimension 6, the only effective operator that is generated at tree level is~\cite{Broncano:2002rw}
\begin{equation}
\delta \mathcal{L}^{d=6} = c^{d=6}_{ij}\left( \left(L_i^\dagger H_u\right)i\slashed{\partial}\left(H_u^\dagger L_j\right) +  \left(L_i^\dagger \tilde{H}_d\right)i\slashed{\partial}\left(\tilde{H}_d^\dagger L_j\right)\right)\,.
\label{eq:d_6op}
\end{equation}
When the Higgs doublets acquire VEVs, $\delta \mathcal{L}^{d=6}$ leads to corrections to the light neutrino kinetic terms, which become non-diagonal. The necessary rotation and normalisation to bring the neutrino kinetic terms to its canonical form induces deviations of unitarity in the leptonic mixing matrix that appears in the charged current (CC) interactions.\\

In the full theory, the renormalisable Yukawa and mass Lagrangians of this minimal model contain the following terms
\begin{eqnarray}
\mathcal{L_\text{Yuk}^\text{MS}}= {y_{i}^\nu H_uL_i\nu^c}+{\epsilon_1y_{i}^{\nu \prime} \tilde H_dL_i\overline{\nu^c}}  \label{eq:L_Yukawa_min} + \text{h.c.} \,,
\label{eq:L_Yukawa_min}
\end{eqnarray}
and
\begin{eqnarray}
\mathcal{L_\text{mass}^\text{MS}}=  {M^{\nu} \nu^c \overline{\nu^c}}+\text{h.c.} \,,
\label{eq:L_mass_min}
\end{eqnarray}
where the transposes in the leptons have been omitted to shorten notation. We assume that the Yukawa couplings between the left-handed neutrinos $\nu_i$,
the vector-like neutrino $\overline{\nu_4^c}$ and $\tilde H_d$ in Eq.~(\ref{eq:L_Yukawa_min}) are suppressed by $\epsilon_1$. 
This assumption allows the Yukawa couplings  between the left-handed neutrinos $\nu_i$, the vector-like neutrino ${\nu_4^c}$ 
and $H_u$ in Eq.~(\ref{eq:L_Yukawa_min}) to be large, leading to possibly observable violations of unitarity.
The key point here is that the effective Weinberg-like operator for neutrino mass involves both the Higgs doublets 
and hence the Yukawa coupling to $H_u$ may be large if that 
to $\tilde H_d$ is small, for a given neutrino mass. 
This is not possible for the usual Weinberg operator arising from the conventional seesaw mechanism, which makes the novel 
seesaw mechanism discussed here interesting.\\

In the following basis, the full neutrino mass matrix reads
\vspace*{-8px}
\begin{equation}
M^\nu=\begin{blockarray}{cccccc}
& \nu_1 & \nu_2 & \nu_3 & \nu^c & \overline{\nu^c} \\
\begin{block}{c(ccccc)}
\nu_1\text{ } & 0 & 0 & 0 & y_{1}^\nu v & \epsilon_1 y_{1}^{\nu\prime} v^\prime \\
\nu_2\text{ } & 0 & 0 & 0 & y_{2}^\nu v & \epsilon_1 y_{2}^{\nu\prime} v^\prime \\
\nu_3\text{ } & 0 & 0 & 0 & y_{3}^\nu v & \epsilon_1 y_{3}^{\nu\prime} v^\prime \\
\nu^c\text{ } & y_{1}^\nu v & y_{2}^\nu v & y_{3}^\nu v & 0 & M^\nu \\
\overline{\nu^c}\text{ } & \epsilon_1 y_{1}^{\nu \prime} v^\prime & \epsilon_1 y_{2}^{\nu\prime} v^\prime & \epsilon_1 y_{3}^{\nu\prime} v^\prime  & M^\nu & 0 \\
\end{block}\\
\end{blockarray}\,\text{ }\, \equiv \left(\begin{array}{cc} 0 & m_D^T  \\ m_D &  M_N\end{array}\right)\,,
\vspace*{-20px}
\label{eq:neu_mass1}
\end{equation}
where $v=v_\text{EW}/\sqrt{2}\simeq 174$ GeV and $v^\prime$ are the VEVs of the Higgs $H_u$ and $H_d$, respectively, and where the Dirac and Majorana mass matrices are defined as
\vspace*{-8px}
\begin{equation}
m_D = \begin{blockarray}{cccc}
& \nu_1 & \nu_2 & \nu_3 \\
\begin{block}{c(ccc)}
 \nu^c \text{ } & y_{1}^\nu v & y_{2}^\nu v  & y_{3}^\nu v   \\
\overline{\nu^c} \text{ } & \epsilon_1 y_{1}^{\nu \prime}v^\prime & \epsilon_1 y_{2}^{\nu\prime} v^\prime & \epsilon_1 y_{3}^{\nu\prime} v^\prime \\
\end{block} \\
\end{blockarray} \quad\quad\quad  \text{and} \quad\quad  
M_N = \begin{blockarray}{ccc}
& \nu^c & \overline{\nu^c} \\
\begin{block}{c(cc)}
 \nu^c \text{  } & 0 & M^\nu    \\
\overline{\nu^c} \text{ } & M^\nu & 0 \\
\end{block} \\
\end{blockarray}\text{  } \text{  } \,.
\vspace*{-20px}
\label{eq:Dirac_Majorana_minimal}
\end{equation}

The neutrino mass matrix of Eq.~(\ref{eq:neu_mass1}) is diagonalised by the full unitary matrix $U$
\begin{equation}
U^T \left(\begin{array}{cc} 0 & m_D^T  \\ m_D &  M_N\end{array}\right) U= \left(\begin{array}{cc} m^\text{diag} & 0  \\ 0 &  M^\text{diag}\end{array}\right)\,,
\label{eq:seesaw}
\end{equation}
where $m^\text{diag}$ and $M^\text{diag}$ are the diagonal matrices containing the masses of the light and heavy sectors, respectively. In all generality, this diagonalisation can be done as the product of two consecutive rotations. This first rotation is a block-diagonalisation, while the second matrix contains the two unitary rotations $V$ and $V^\prime$ that diagonalise the masses of the light and heavy neutrinos, respectively. Since the rotation between the two heavy states is unphysical, $V^\prime=I$ can be used, and thus, the full unitary neutrino mixing matrix $U$ is given by
\begin{equation}
U= \left(\begin{array}{cc} A_{11} & A_{12}  \\ A_{21} &  A_{22}\end{array}\right)\left(\begin{array}{cc} V & 0  \\ 0 &  I \end{array}\right)\,,
\label{eq:U_tot}
\end{equation}
where the block-diagonalisation can be parametrise as the exponential of a block off-diagonal anti-Hermitian complex matrix $\Theta$~\cite{Blennow:2011vn}
\begin{equation}
\left(\begin{array}{cc} A_{11} & A_{12}  \\ A_{21} &  A_{22}\end{array}\right)= \exp\left(\begin{array}{cc} 0 & \Theta  \\ -\Theta^\dagger &  0\end{array}\right)= \left(
\begin{array}{cc}
\displaystyle\sum\limits_{n=0}^\infty \frac{ \left(- \Theta \Theta^\dagger \right)^{n}}{(2n)!} & 
\displaystyle\sum\limits_{n=0}^\infty \frac{ \left(- \Theta \Theta^\dagger \right)^{n}}{\left(2n+1\right)!} \Theta  \\ 
-\displaystyle\sum\limits_{n=0}^\infty \frac{ \left(- \Theta^\dagger \Theta \right)^{n}}{\left(2n+1\right)!} \Theta^\dagger & 
\displaystyle\sum\limits_{n=0}^\infty \frac{ \left(- \Theta^\dagger \Theta \right)^{n}}{2n!}\end{array}
\right)\,.
\end{equation}

When substituting Eq.~(\ref{eq:U_tot}) in Eq.~(\ref{eq:seesaw}), and considering that the mass scale of the vector-like neutrinos $M_4^\nu$ is much higher than the VEVs $v$ and $v^\prime$, i.e. $m_D\ll M_N$, the usual seesaw relations are recovered
\begin{eqnarray}
\nonumber
\Theta &\simeq& m_D^\dagger M_N^{-1} \,,\\
V^* m^\text{diag} V^\dagger  &\simeq& - m_D^T M_N^{-1} m_D \equiv -\hat{m}\,,\\
M^\text{diag} &\simeq& M_N \nonumber \,, 
\label{eq:seesaw_relations}
\end{eqnarray}
with $\hat{m}\equiv -v v^\prime c^{d=5}$ the coefficient of the dim-5 new type of Weinberg operator that generates the light neutrino masses of Eq.~(\ref{eq:d_5op}). Therefore $V$ is approximately the unitary rotation that diagonalises the light neutrinos, and can be identified as $U_\text{PMNS}$, the Pontecorvo-Maki-Nakagawa-Sakata (PMNS) mixing matrix measured in neutrino oscillation experiments and parametrised~\cite{Chau:1984fp} as $U_\text{PMNS}= U_{23}\left(\theta_{23}\right) U_{13}\left(\theta_{13},\delta\right) U_{12}\left(\theta_{12}\right) \text{diag}\left(e^{-i \alpha^\prime/2}, e^{-i \alpha/2},1\right)$\footnote{In the minimal scenario, $\alpha^\prime=0$.}. At leading order in $\Theta$, the full mixing matrix $U$ will be
\begin{equation}
U\simeq \left(\begin{array}{cc}I-\dfrac{\Theta\Theta^\dagger}{2} & \Theta  \\ -\Theta^\dagger  &  I-\dfrac{\Theta\Theta^\dagger}{2}\end{array}\right)\left(\begin{array}{cc} U_\text{PMNS} & 0 \\ 0 & I \end{array} \right)\,,
\label{eq:final_U}
\end{equation}
where its first sub-block parametrises the mixing of the light sector~\cite{FernandezMartinez:2007ms}
\begin{equation}
N\equiv \left(I-\dfrac{\Theta\Theta^\dagger}{2}\right)U_\text{PMNS} = \left(I-\eta\right)U_\text{PMNS}\,.
\end{equation}
Thus, the presence of the heavy vector-like family induces non-unitarity in the mixing matrix that appear in the charged current interactions. These deviations of unitarity of the leptonic mixing matrix induced by the dim-6 operator of Eq.~(\ref{eq:d_6op}), are parametrised by the hermitian matrix $\eta\equiv v^2 c^{d=6}/2$
\begin{equation}
\eta = \dfrac{\Theta \Theta^\dagger}{2} = \dfrac{1}{2}m_D^\dagger M_N^{-2} m_D\,.
\label{eq:dim6}
\end{equation}

In terms of the Yukawa couplings, the light neutrino mass matrix of Eq.~(\ref{eq:seesaw_relations}) built up from the Dirac and Majorana mass matrices of Eq.~(\ref{eq:Dirac_Majorana_minimal}) reads
\begin{equation}
\hat{m}_{ij}=\dfrac{ \epsilon_1 v  v^\prime}{M^\nu}\left(y_{i}^\nu y_{j}^{\nu\prime} + y_{i}^{\nu\prime} y_{j}^\nu\right)\,,
\label{eq:dim5_relation_simp}
\end{equation} 
where it can be seen that the smallness of the light neutrino masses stem not only from the suppression of $M^\nu$, but also from the small size of $\epsilon_1$. On the other hand, the deviations of unitarity will be
\begin{equation}
\eta_{ij}=\dfrac{1}{2 M^{\nu 2}}\left(v^2 y_{i}^{\nu *} y_{j}^\nu +\epsilon_1^2 v^{\prime 2} y_{i}^{\nu \prime *} y_{j}^{\nu \prime}\right)\simeq \dfrac{v^2}{2 M^{\nu 2}}y_{i}^{\nu*} y_{j}^\nu\,,
\label{eq:dim6_relation_simp}
\end{equation}
where the second term can be safely neglected since it would be of the order of the neutrino mass scale squared over $v^2$. Therefore, in this model the deviations of unitarity of the PMNS matrix are not suppressed by $\epsilon_1$, and could be arbitrarily large. At leading order, the deviations of unitarity are thus determined only by the first row of $m_D$ containing the 3 complex Yukawa couplings $y_{i}^\nu$, and the mass scale of the vector-like neutrino $M^{\nu}$.\\

However, since both $\eta$ and $\hat m$ are built from $m_D$ and $M_N$, they may not be fully independent. This implies that in determinate cases, $\eta$ could be partially  reconstructed from $\hat m$, and therefore, from the observed pattern of neutrino masses and mixings in neutrino oscillation experiments. In the particular case of this minimal scenario, the Yukawa couplings $y_{i4}^\nu$ ($y_{i4}^{\nu \prime}$) of Eq.~(\ref{eq:Dirac_Majorana_minimal}) will be determined~\cite{Gavela:2009cd} up to an overall factor $y$ ($y^\prime$) from the elements of the PMNS mixing matrix, and the two mass squared splittings, $\Delta m_\text{sol}^2$ and $\Delta m_\text{atm}^2$. 
Notice that in this minimal scenario just two light neutrinos get masses, and that therefore, the lightest neutrino is strictly massless\footnote{The lightest neutrino is still massless when the 1-loop neutrino mass corrections that arise from the neutrino self-energy are considered~\cite{Adhikari:2010yt}.}. On the other hand, since the hierarchy of the neutrinos is not determined yet, there will be two possible relations for the Yukawa couplings. For a normal hierarchy (NH), $m_1=0$ and the Yukawa couplings read
\begin{eqnarray}
y_{i}^\nu&=&\dfrac{y}{\sqrt{2}}\left( \sqrt{1+\rho} \left(U_\text{PMNS}^*\right)_{i3}+\sqrt{1-\rho} \left(U_\text{PMNS}^*\right)_{i2}\right)\,, \label{eq:Yukawa_relation_NH}\\
y_{i}^{\nu \prime}&=&\dfrac{y^\prime}{\sqrt{2}}\left( \sqrt{1+\rho} \left(U_\text{PMNS}^*\right)_{i3}-\sqrt{1-\rho} \left(U_\text{PMNS}^*\right)_{i2}\right)\,, \nonumber
\end{eqnarray}
where $y$ and $y^\prime$ are real numbers, and where $\rho=(1-\sqrt{r})/(1+\sqrt{r})$ with $r\equiv\vert \Delta m_\text{sol}^2\vert/\vert \Delta m_\text{atm}^2\vert = \Delta m_{21}^2/\Delta m_{31}^2$. While for an inverted hierarchy (IH), $m_3=0$ and the Yukawa couplings are given by
\begin{eqnarray}
y_{i}^\nu&=&\dfrac{y}{\sqrt{2}}\left( \sqrt{1+\rho} \left(U_\text{PMNS}^*\right)_{i2}+\sqrt{1-\rho} \left(U_\text{PMNS}^*\right)_{i1}\right)\,, \label{eq:Yukawa_relation_IH} \\
y_{i}^{\nu \prime}&=&\dfrac{y^\prime}{\sqrt{2}}\left( \sqrt{1+\rho} \left(U_\text{PMNS}^*\right)_{i2}-\sqrt{1-\rho} \left(U_\text{PMNS}^*\right)_{i1}\right)\,,  \nonumber
\end{eqnarray}
where now $\rho=(1-\sqrt{1+r})/(1+\sqrt{1+r})$ with $r = \Delta m_{21}^2/\Delta m_{32}^2$. As a result, all the neutrino phenomenology of this minimal scenario is described by five free parameters: two real numbers $y$ and $y^\prime$, two phases $\delta$ and $\alpha$, and one mass scale $M_4^\nu$. But only four of them will enter in the description of the deviations of unitarity through Eq.~(\ref{eq:dim6_relation_simp}). \\

\begin{figure}
\centering
\includegraphics[width=0.4\textwidth]{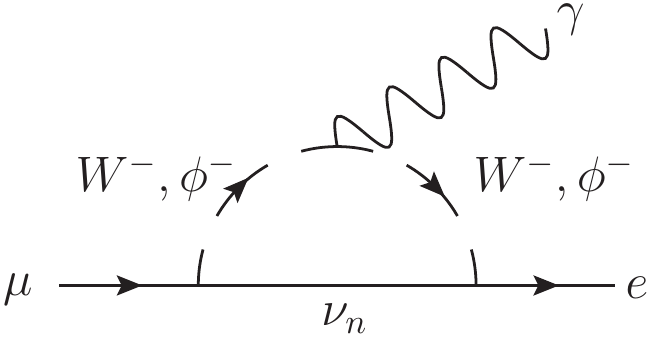}
\caption{Diagram showing the extra neutrino contributions to $\mu\to e \gamma$. Here $\nu_n$ refers to the neutrinos in the mass basis, and $\phi^-$ represents the Goldstone boson.}
\label{fig:meg}
\end{figure}

Since the presence of the extra heavy vector-like neutrinos induces deviation of unitarity in the PMNS matrix, the GIM cancellation~\cite{Glashow:1970gm} that suppresses flavour-changing processes is loss. As a result, the present limits on LFV processes will set a strong constrain on the non-unitarity of the leptonic mixing matrix, and therefore on the free parameters of the minimal scenario $y$, $\delta$ and $\alpha$ through Eq.~(\ref{eq:dim6_relation_simp}). In particular, the nowadays strongest constrain on the elements of the $\eta$ matrix comes from $\mu\to e\gamma$. Figure~(\ref{fig:meg}) shows the extra contribution to the radiative decay $\mu\to e\gamma$ in presence of the vector-like neutrinos of the model. 

The contribution to the branching ratio from both the heavy vector-like neutrinos and the light neutrinos $\nu_i$ is given by 
\begin{equation}
\dfrac{\Gamma\left(\mu\to e \gamma \right)}{\Gamma\left(\mu\to e \nu_\mu \overline{\nu}_e \right)}=\dfrac{3\alpha}{32\pi}\dfrac{\vert \displaystyle\sum_{n=1}^5 U_{2 n} U^\dagger_{n1} F(x_n)\vert^2}{\left(U U^\dagger\right)_{11}\left(U U^\dagger\right)_{22}}\,,
\label{eq:meg}
\end{equation}
where $x_n=M_n^2/M_W^2$, and where $F(x_n)$ reads
\begin{equation}
F(x_n) = \dfrac{10-43x_n+78x_n^2-\left(49-18 \log x_n\right) x_n^3+4x_n^4}{3\left(x_n-1\right)^4}\,,
\end{equation}
For masses of the vector-like neutrinos $M^\nu\gg M_W$, the sum in Eq.~(\ref{eq:meg}) can be separated in light and heavy sectors factorizing the corresponding $F(x_n)$ function
 \begin{equation}
\dfrac{\Gamma\left(\mu\to e \gamma \right)}{\Gamma\left(\mu\to e \nu_\mu \overline{\nu}_e \right)}\simeq\dfrac{3\alpha}{8\pi}\vert \eta_{21}\vert^2 \left(F(\infty) -F(0)\right)^2= \dfrac{3\alpha}{2\pi}\vert \eta_{21}\vert^2 \,,
\label{eq:meg_simp}
\end{equation}
where can be seen that loss of the GIM cancellation comes from the difference of the two mass scales involved, and the non-unitarity of the leptonic mixing matrix. When comparing with the existing present experimental limit~\cite{Tanabashi:2018oca} of the radiative decay, the following upper bound at $1\sigma$ is derived~\cite{Fernandez-Martinez:2016lgt}
\begin{equation}
\vert\eta_{21}\vert\leq 8.4 \cdot 10^{-6} \,.
\label{eq:eta_limit}
\end{equation} 

\begin{figure}
\centering
\includegraphics[width=0.6\textwidth]{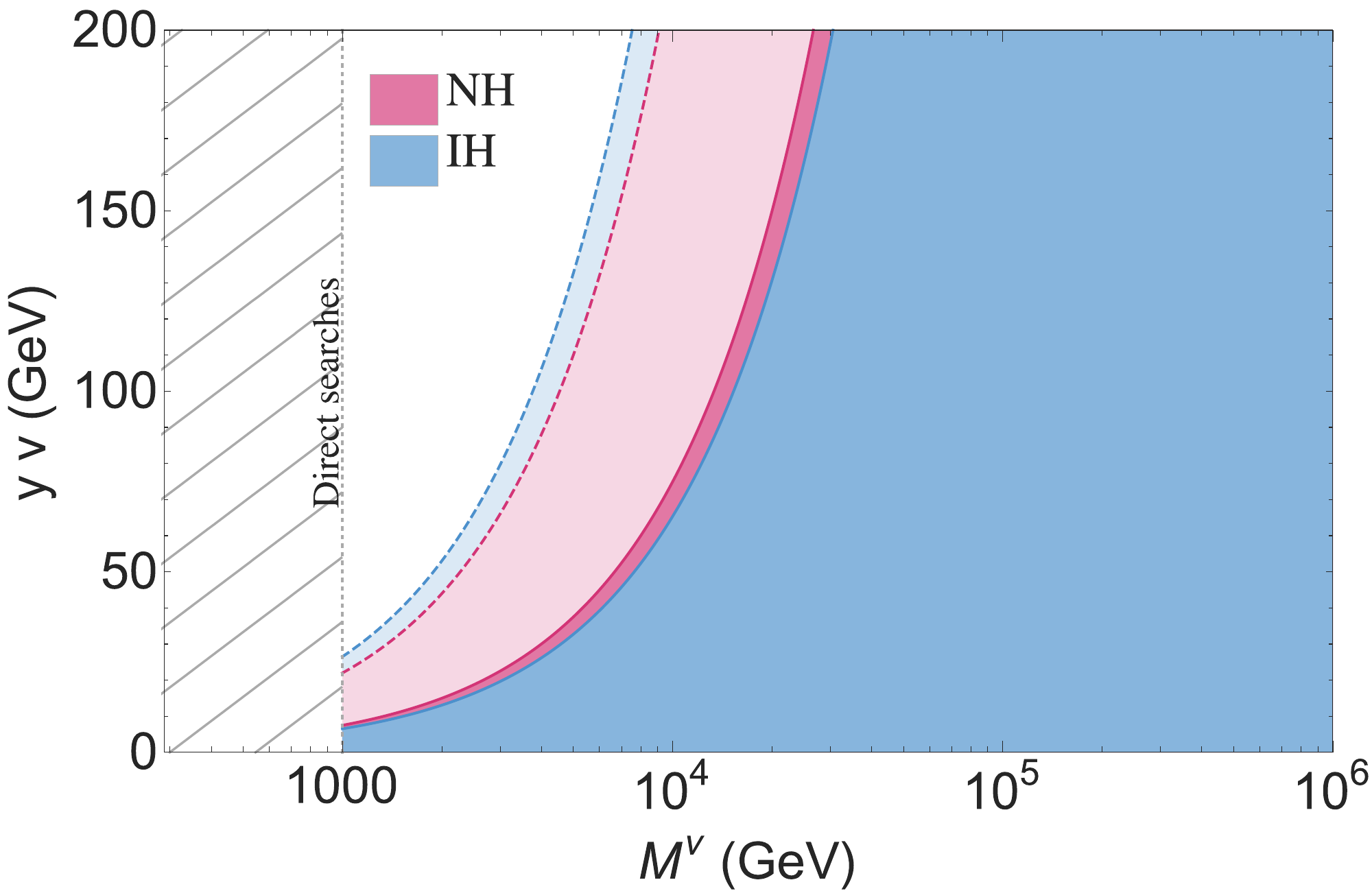}
\caption{Allowed region of the free parameters $y$ and $M^\nu$ in the minimal scenario when the present bound~\cite{Tanabashi:2018oca} on $\mu\to e\gamma$ is considered. For each hierarchy, the boundary ranges from the solid to the dashed line depending on the values of the phases $\delta$ and $\alpha$. The pink region corresponds to NH while the blue region corresponds to IH. The hatched gray area has been excluded by direct searches~\cite{Aad:2015kqa}.}
\label{fig:Yuk_limit} 
\end{figure}

In Figure~\ref{fig:Yuk_limit} the allowed region of the free parameters of the minimal scenario is shown. The hatched gray region is excluded by direct searches in ATLAS~\cite{Aad:2015kqa}, while the pink (blue) regions correspond to the allowed values of $y$ and $M_4^\nu$ when the present constrain on $\eta_{12}$ of Eq.~(\ref{eq:eta_limit}) and a NH (IH) in the light neutrino sector is considered. The allowed region depends on the CP-violating phase $\delta$ and the Majorana phase $\alpha$ of the PMNS matrix. The boundaries of the allowed regions depend on the values of the free phases $\delta$ and $\alpha$. For a NH (IH), the solid line correspond to $\delta=\alpha=0$ ($\delta\simeq\pi/2$, $\alpha\simeq9\pi/10$) and can be relax till the dashed line which corresponds to $\delta=0$ and $\alpha=2 \pi$ ($\delta=\alpha=0$). For the numerical analysis, the central values of the $\theta_{ij}$ mixing angles of the PMNS matrix, the solar and the atmospheric mass splittings of the NuFIT 4.0~\cite{Esteban:2018azc} have been adopted.\\

The $U(1)^\prime$ charge of the two Higgs doublets forbids the usual Yukawa couplings for the charged fermions $y^\psi_{ij}H\psi_i\psi_j^c$. However, if one power of the scalar $\phi$ is introduced, the $U(1)^\prime$ charge would be absorbed, and non-renormalisable Yukawa operators of the form $y^\psi_{ij}H\psi_i\phi\psi_j^c$ would be allowed. In order to build a renormalisable model, we will enlarge the particle content of this simplify model by a fourth vector-like family that will allow to generate masses for all the charged fermions via effective Yukawa couplings, as proposed in Ref.~\cite{King:2018fcg}. 

%%%%%%%%%%%%%%%%%%%%%%%
\section{Renormalisable Type Ib (plus Type Ia) Seesaw Model}
\label{s:general_model}
%%%%%%%%%%%%%%%%%%%%%%%

The model of the previous section does not allow renormalisable Yukawa couplings for the charged fermions and so must be extended somehow.
Here we identify the two right-handed neutrinos as originating from a fourth vector-like family, whose presence also allows
for the generation of effective Yukawa couplings. Notice that the vector-like structure makes the model anomaly-free since the anomalies cancel between conjugate representations in the fourth family~\cite{King:2017anf}.

The particle content of the general model that we consider here consists in three left-handed families $\psi_i=Q_i, L_i$, the CP conjugated right handed families $\psi_i^c = u_i^c, d_i^c, e_i^c$ (excluding the right-handed neutrinos)
and a fourth vector-like left-handed family consisting in 
$\psi_4=Q_4, L_4$, and $\psi_4^c= u_4^c, d_4^c, e_4^c, \nu_4^c$ and the conjugate representations 
$\overline{\psi_4}=\overline{Q_4}, \overline{L_4}$, and 
$\overline{\psi_4^c}= \overline{u_4^c}, \overline{d_4^c}, \overline{e_4^c}, \overline{\nu_4^c}$. 
Here we identify $\nu_4^c$ and $\overline{\nu_4^c}$ with $\nu^c$ and $\overline{\nu^c}$ of the minimal type Ib seesaw model
of the previous section.
So far we have not included any genuine right-handed neutrino $N^c$ (neutral under $U(1)'$).
However, later in this section we shall consider the additional 
effect of including (in addition to the fourth family states) 
one CP conjugated right-handed singlet neutrino $N^c$ in the seesaw mechanism.
Notice that here $\overline{\psi}$ denotes that the fermion is in the conjugate representation of the SM gauge group. 
In our notation all these fermion fields $\psi_i,\psi_i^c,\psi_4,\psi_4^c, \overline{\psi_4}, \overline{\psi_4^c}$ transform as left-handed spinors under the Lorentz group. 
The vector-like family is charged under a gauge symmetry $U(1)^\prime$ with charges $+1$ ($-1$) for $\psi_4,\psi_4^c$ ($\overline{\psi_4},\overline{\psi_4^c}$). However, since the model is ``fermiophobic", the three chiral families 
$\psi_i,\psi_i^c$ are neutral under the $U(1)^\prime$ symmetry. The singlet scalar field $\phi$ is the responsible of breaking the $U(1)^\prime$ symmetry developing vacuum expectation value (VEV) $\left\langle \phi\right\rangle$ around the TeV scale. The $Z^\prime$ boson generated after the symmetry breaking has a mass at the same scale. The scalar $\phi$ has $U(1)^\prime$ charge $+1$. Since the two Higgs doublets  $H_u$ and $H_d$ are negatively charged under the $U(1)^\prime$, no standard renormalisable Yukawa couplings among the first three chiral families are allowed, and only those which couple the first three chiral families to the fourth vector-like family are generated. All the charges of the different left-handed particles of the model are summarised in Table~\ref{t:Particle_content2}.\\

The renormalisable Yukawa and mass Lagrangians that account for the interactions of the particles summarised in Table~\ref{t:Particle_content2} are
\begin{eqnarray}
\mathcal{L_\text{Yuk}}&=& y_{i4}^\psi H \psi_i \psi_4^c + y_{i4}^{\psi\prime} H^* \psi_i \overline{\psi_4^c} + y_{4i}^\psi H \psi_4\psi_i^c + \text{h.c.}\, ,
\label{eq:L_Yuk} \\
\mathcal{L_\text{mass}}&=& x_i^\psi \phi \psi_i\overline{\psi_4}+ x_i^{\psi^c} \phi \psi_i^c\overline{\psi_4^c}+M_4^\psi \psi_4\overline{\psi_4} + M_4^{\psi^c}\psi_4^c\overline{\psi_4^c}+\text{h.c.}\,,
\label{eq:L_mass}
\end{eqnarray} 
where $x_i^\psi$ and $y_{ij}^\psi$ are dimensionless coupling constants and $M_4^\psi$ are explicit mass terms.\\

Notice that the two Higgs doublets $H$ are charged under $U(1)^\prime$, and thus the usual Yukawa couplings $y_{ij}^\psi H\psi_i\psi_j^c$ are forbidden for $i, j = 1,...,3$. However, effective $3\times3$ Yukawa couplings may be generated by the two mass insertion diagrams shown Fig.~\ref{fig:Eff_Yukawa}. These effective Yukawa couplings read
\begin{equation}
\mathcal{L}^\text{eff}_\text{Yuk}=\left(\dfrac{y_{i4}^\psi x_j^{\psi^c}\left\langle\phi\right\rangle}{M_4^{\psi^c}}+\dfrac{ x_i^{\psi} y_{4j}^\psi\left\langle\phi\right\rangle}{M_4^{\psi}}\right)H\psi_i\psi_j^c+\text{h.c.}\,.
\label{eq:eff_Yukawa}
\end{equation}
\begin{figure}[htp]
\centering
\includegraphics[width=0.4\textwidth]{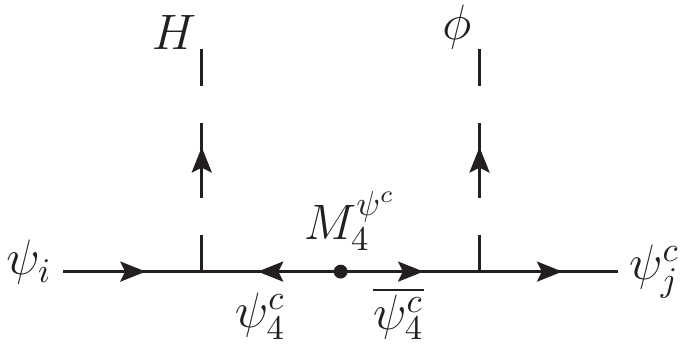} \quad\quad\quad
\includegraphics[width=0.4\textwidth]{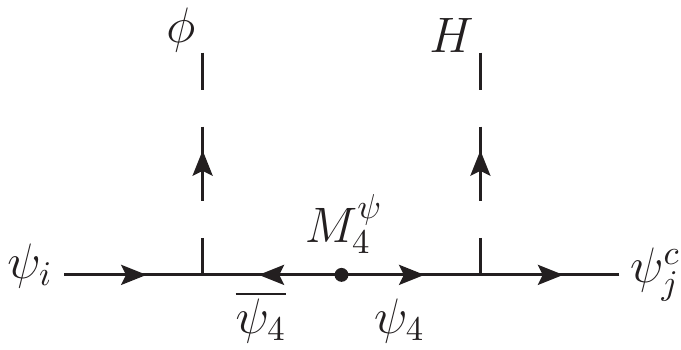}
\caption{Mass insertion approximation diagrams which lead to the effective Yukawa couplings. $H$ represents the two Higgs doublets $H_{u,d}$.}
\label{fig:Eff_Yukawa}
\end{figure}

\begin{table}[htp]
\begin{center}
\begin{tabular}{|c||ccc|c|}
\hline 
Field &  $ SU(3)_c $  &  $ SU(2)_L $  & $U(1)_Y$ & $U(1)^\prime$ \\
\hline 
\hline
$Q_i$ & \textbf{3} & \textbf{2} & $1/6$ & 0 \\
$u_i^c$ & $\mathbf{\overline{3}}$ &  \textbf{1} &  $-2/3$ &  0 \\
 $d_i^c$ &  $\mathbf{\overline{3}}$ &  \textbf{1} &  $1/3$ &  0 \\
 $L_i$ &  \textbf{1} &  \textbf{2} &  $-1/2$ &  0 \\
 $e_i^c$ &  \textbf{1} &  \textbf{1} &  1 &  0 \\
 \hline
 \hline
 $N^c$ &  \textbf{1} &  \textbf{1} &  0 &  0 \\
\hline
\hline
$Q_4$ &  \textbf{3} &  \textbf{2} &  $1/6$  & 1  \\
$u_4^c$ &  $\mathbf{\overline{3}}$ & \textbf{1} & $-2/3$ & 1 \\
$d_4^c$ &  $\mathbf{\overline{3}}$ & \textbf{1} & $1/3$ & 1 \\
$L_4$ & \textbf{1} & \textbf{2} & $-1/2$ & 1 \\
$e_4^c$ & \textbf{1} & \textbf{1} & $1$ & 1 \\
$\nu_4^c$ & \textbf{1} & \textbf{1} & 0 & 1 \\
\hline
\hline
$\overline{Q_4}$ & $\mathbf{\overline{3}}$ & $\mathbf{\overline{2}}$ & $-1/6$ & $-1$ \\
$\overline{u_4^c}$ & \textbf{3} & \textbf{1} & $2/3$ & $-1$ \\
$\overline{u_4^c}$ & \textbf{3} & \textbf{1} & $-1/3$ & $-1$ \\
$\overline{L_4}$ & \textbf{1} & $\mathbf{\overline{2}}$ & $1/2$ & $-1$ \\
$\overline{e_4^c}$ & \textbf{1} & \textbf{1} & $-1$ & $-1$ \\
$\overline{\nu_4^c}$ & \textbf{1} & \textbf{1} & 0 & $-1$ \\
\hline
\hline
 $\phi$ &  \textbf{1} &  \textbf{1} &  0 &  1 \\
\hline
\hline
 $H_u$ &  \textbf{1} &  \textbf{2} &  $1/2$ &  $-1$ \\
 $H_d$ &  \textbf{1} &  \textbf{2} &  $-1/2$ &  $-1$ \\
\hline
\end{tabular}
\end{center}
\caption{The most general model consists of three left-handed families $\psi_i=Q_i, L_i$ and its CP conjugated right-handed fields $\psi^c_i=u_i^c, d_i^c, e_i^c$ $(i=1,2,3)$, and a fourth vector-like family consisting of $\psi_4$ plus $\overline{\psi_4}$ and $\psi_4^c$ plus $\overline{\psi_4^c}$, together with the $U(1)^\prime$-breaking scalar field $\phi$ and the two Higgs scalar doublets $H_u$ and $H_d$ which are charged under $U(1)^\prime$. In the minimal model, the 
single CP conjugated right-handed neutrino $N^c$ is not introduced, and will be considered only later. Notice that all the fermions of this table are left-handed spinors and the bars indicate conjugate representations under the SM gauge group.} 
\label{t:Particle_content2}
\end{table}

In the minimal scenario we did not consider a full vector-like fourth family. Now including such states, the general scenario (GS) also involves one CP conjugate heavy right-handed neutrino $N^c$ as summarised in Table~\ref{t:Particle_content2}. This $N^c$ is singlet under all the gauge group and therefore a Majorana mass $M_M$ is allowed for it. The Yukawa and mass Lagrangians of the general scenario will now contain the following terms
\begin{eqnarray}
\mathcal{L_\text{Yuk}^\text{GS}}&=& y_{i4}^u H_uQ_iu_4^c + y_{i4}^d H_dQ_id_4^c+\boldsymbol{y_{i4}^\nu H_uL_i\nu_4^c}+\boldsymbol{\epsilon_1y_{i4}^{\nu \prime} \tilde H_d L_i\overline{\nu_4^c}} +y_{4i}^eH_d L_i e_4^c \label{eq:L_Yukawa_gen} \\
	&+& y_{4i}^uH_uQ_4u_i^c + y_{4i}^dH_dQ_4d_i^c+y_{4i}^eH_d L_4 e_i^c +  \boldsymbol{\epsilon_2 y_{4}^N H_u L_4 N^c} + \text{h.c.}  \nonumber\,,
\end{eqnarray}
and
\begin{eqnarray}
\mathcal{L_\text{mass}^\text{MS}}&=& x_i^Q\phi Q_i \overline{Q_4}+x_i^L\phi L_i \overline{L_4}+x_i^{u}\phi u_i^c \overline{u_4^c}+x_i^{d}\phi d_i^c \overline{d_4^c}+ x_i^{e}\phi e_i^c \overline{e_4^c} +   \boldsymbol{x^{N} \phi N^c \overline{\nu_4^c}} +  \boldsymbol{x^{N \prime} \phi^c N^c \nu_4^c} \label{eq:L_mass_gen}\\
&+& M_4^Q Q_4 \overline{Q_4}+ M_4^L L_4 \overline{L_4}+M_4^{u} u_4^c \overline{u_4^c}+M_4^{d} d_4^c \overline{d_4^c}+ M_4^{e} e_4^c \overline{e_4^c}+ \boldsymbol{M_4^{\nu} \nu_4^c \overline{\nu_4^c}} + \boldsymbol{\dfrac{1}{2} M_M N^c N^c} +\text{h.c.}  \,,\nonumber
\end{eqnarray}
where again we are supposing that the Yukawa coupling between the vector-like lepton $L_4$ and the heavy neutrino $N^c$ is suppressed by $\epsilon_2$. Moreover, we will consider that both $\mu_4\equiv x^{N} \langle \phi \rangle$ and $\mu_3\equiv x^{N \prime} \langle \phi \rangle$ are suppressed compared to the Majorana scale (i.e. $\mu_{3,4}\ll M_M$).
The terms of the Lagrangians of Eq.~(\ref{eq:L_Yukawa_gen}) and Eq.~(\ref{eq:L_mass_gen}) that will enter in the discussion of this section are highlighted in bold face and may be compared to the corresponding terms in the minimal type Ib seesaw
model in Eqs.~(\ref{eq:L_Yukawa_min}) and (\ref{eq:L_mass_min}), where we identify 
$\nu_4^c$ and $\overline{\nu_4^c}$ with $\nu^c$ and $\overline{\nu^c}$ and the parameters 
$y_{i4}^{\nu} , y_{i4}^{\nu \prime}, M_4^{\nu}$ with $y_{i}^\nu , y_{i}^{\nu \prime}, M^{\nu}$.

\subsubsection{The effective Yukawa couplings}

As explained at the beginning of this section, the presence of $N^c$ allows to generate the effective Yukawa interaction $H_u L_i N^c$ of Eq.~(\ref{eq:eff_Yukawa}) through the diagrams in the mass insertion approximation shown in Figure~\ref{fig:Eff_Yukawa}.
 
\begin{figure}[htp]
\centering
\includegraphics[width=0.4\textwidth]{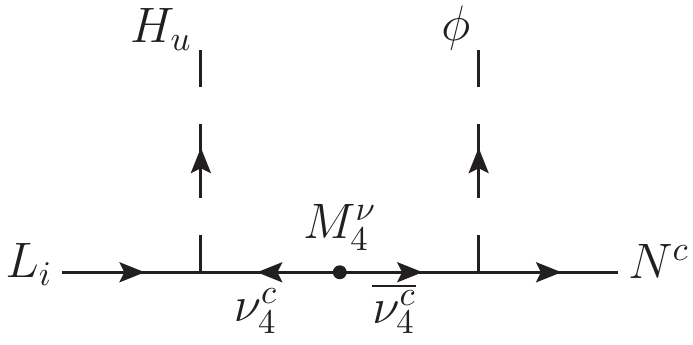}\quad\quad\quad
\includegraphics[width=0.4\textwidth]{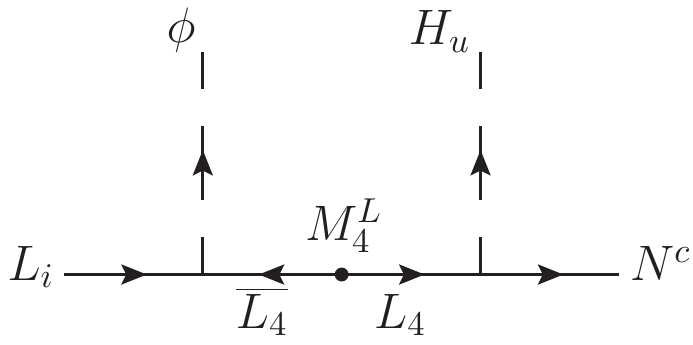}
\caption{Diagrams in the mass insertion approximation that generate the effective Yukawa couplings that will contribute to the light neutrino masses.}
\label{fig:Eff_Yukawa}
\end{figure}

As a result, this effective Yukawa interaction will generate the following effective Dirac masses
\begin{equation}
m_{D_i}^\text{eff}=\dfrac{\mu_3}{M_4^\nu} y_{i4}^{\nu^c} v +\dfrac{x_i^{L}\langle \phi \rangle}{M_4^L} \epsilon_2y_{4}^{N} v \,.
\label{eq:effective_Dirac_mass}
\end{equation}

\subsubsection{Generating neutrino masses}

The full neutrino mass matrix of the general scenario will be given by the right-hand side term of Eq.~(\ref{eq:neu_mass1}), where now the Dirac and Majorana mass matrices read
\vspace*{-8px}
\begin{equation}
m_D = \begin{blockarray}{cccc}
& \nu_1 & \nu_2 & \nu_3 \\
\begin{block}{c(ccc)}
 \nu_4^c \text{ } & y_{14}^\nu v & y_{24}^\nu v  & y_{34}^\nu v   \\
\overline{\nu_4^c} \text{ } & \epsilon_1 y_{14}^{\nu \prime}v^\prime & \epsilon_1 y_{24}^{\nu\prime} v^\prime & \epsilon_1 y_{34}^{\nu\prime} v^\prime \\
N^c \text{ } & m_{D_1}^\text{eff} & m_{D_2}^\text{eff} & m_{D_3}^\text{eff} \\
\end{block} \\
\end{blockarray} \quad\quad\quad  \text{and} \quad\quad  
M_N = \begin{blockarray}{cccc}
& \nu_4^c & \overline{\nu_4^c} & N^c \\
\begin{block}{c(ccc)}
 \nu_4^c \text{ } & 0 & M_4^\nu  & \mu_3   \\
\overline{\nu_4^c} \text{ } & M_4^\nu & 0 & \mu_4 \\
N^c \text{ } & \mu_3 & \mu_4 & M_M \\
\end{block} \\
\end{blockarray} \text{  } \text{  } \,.
\vspace*{-20px}
\label{eq:Dirac_Majorana_extra}
\end{equation}

Substituting these Dirac and Majorana mass matrices of Eq.~(\ref{eq:Dirac_Majorana_extra}) into Eq.~(\ref{eq:seesaw_relations}) and assuming that $\epsilon_1 v$, $\epsilon_2 v$, $\mu_3$, and $\mu_4\ll M_4^\nu$ and  $M_M$, the light neutrino mass matrix $\hat{m}$ will be given\footnote{At leading order in the small parameters $\epsilon_1$, $\epsilon_2$, and $\mu_{3,4}/M_X$, with $M_X=M_4^\nu,M_M.$} by
\begin{eqnarray}
\hat{m}_{ij}&\simeq&\dfrac{ \epsilon_1 v  v^\prime}{M_4^\nu}\left(y_{i4}^\nu y_{j4}^{\nu\prime} + y_{i4}^{\nu\prime} y_{j4}^\nu\right)+\dfrac{v^2 \left(\mu_3 -\mu_4\right)^2}{M_M M_4^{\nu 2}}y_{i4}^\nu y_{j4}^\nu +\dfrac{\epsilon_2^2 v^2 \langle \phi \rangle^2}{M_M M_4^{\nu 2}}x_i^L x_j^L y_4^{N 2}  \label{eq:dim5_relation_extra}\\ 
&+&\dfrac{\epsilon_2 v^2 \langle \phi \rangle  \left(\mu_3 -\mu_4\right)y_{4}^N}{M_M M_4^{\nu 2}}\left(x_i^L y_{j4}^\nu+x_j^L y_{i4}^\nu\right)\,. \nonumber
\end{eqnarray}

Without assuming fine-tuning cancellations, all the terms of Eq.~(\ref{eq:dim5_relation_extra}) have to be of the order of the scale of light neutrino masses. And therefore, when computing the $\eta$ matrix in the general scenario, the contributions from the second and third rows of $m_D$ are found to be negligible. As a result, the same relation for $\eta$ of the minimal scenario given by Eq.~(\ref{eq:dim6_relation_simp}) will be recovered
\begin{equation}
\eta_{ij}\simeq \dfrac{v^2}{2 M_4^{\nu 2}}y_{i4}^{\nu*} y_{j4}^\nu\,.
\label{eq:dim6_relation_general}
\end{equation}

Once again, in order to get the correct structure of the symmetric matrix $\hat{m}$, there will be correlations among the elements of $m_D$. As a result, one of the three complex Yukawa couplings $y_{i4}^\nu$ necessary to describe the deviations of unitarity will be completely determined from the other two Yukawa couplings and the elements of $\hat{m}$ as follows~\cite{Fernandez-Martinez:2015hxa}
\begin{eqnarray}
y_{34}^\nu& \simeq& \frac{1}{\hat{m}_{12}^2 - \hat{m}_{11} \hat{m}_{22}}\Bigg( y_{14}^\nu\left(\hat{m}_{12}\hat{m}_{23}-\hat{m}_{13}\hat{m}_{22}\right) \nonumber \\
&+&y_{24}^\nu\left(\hat{m}_{12}\hat{m}_{13}-\hat{m}_{11}\hat{m}_{23}\right)\pm\sqrt{y_{14}^{\nu 2}\hat{m}_{22}-2y_{14}^\nu y_{24}^\nu \hat{m}_{12}+y_{34}^{\nu 2}\hat{m}_{11}}\times  \label{eq:Y3_relation}\\
&\times& \sqrt{\hat{m}_{13}^2\hat{m}_{22}-2\hat{m}_{12}\hat{m}_{13}\hat{m}_{23}+\hat{m}_{11}\hat{m}_{23}^2+\hat{m}_{12}^2\hat{m}_{33}-\hat{m}_{11}\hat{m}_{22}\hat{m}_{33}}\Bigg)\nonumber \,.
\end{eqnarray}

Therefore, of the $\eta$ matrix will be described by: two complex Yukawa couplings $y_{14}^\nu$ and $y_{24}^\nu$ one heavy vector-like neutrino mass scale $M_4^{\nu}$, and the four yet unknown parameters on $\hat m$: the light neutrino mass scale $m_1,3$ and the three phases of the PMNS matrix $\delta$, $\alpha$,$\alpha^\prime$.\\

Similarly, the allowed region of the free parameters $y_{14}^\nu$, $y_{24}^\nu$ and $M_4^{\nu}$ can be analysed by using the present bounds on the elements of $\eta$ through Eq.~(\ref{eq:dim6_relation_general}). In particular, the global-fit to Electroweak and flavour precision observables performed in~\cite{Fernandez-Martinez:2016lgt} sets the following upper bounds
\begin{equation} 
\text{NH} \left\lbrace \begin{array}{c}
\eta_{11} < 4.2\cdot 10^{-4}  \\
\eta_{22} < 2.9\cdot 10^{-7}
\end{array} \right.
\quad \text{and} \quad
\text{IH} \left\lbrace \begin{array}{c}
\eta_{11} < 4.8 \cdot 10^{-4}  \\
\eta_{22} < 2.4 \cdot 10^{-7}
\end{array} \right. \,,
\label{eq:eta_bounds}
\end{equation} 
at $1\sigma$ for both normal and inverted hierarchies.\\

\begin{figure}[htp]
\centering
\includegraphics[width=0.6\textwidth]{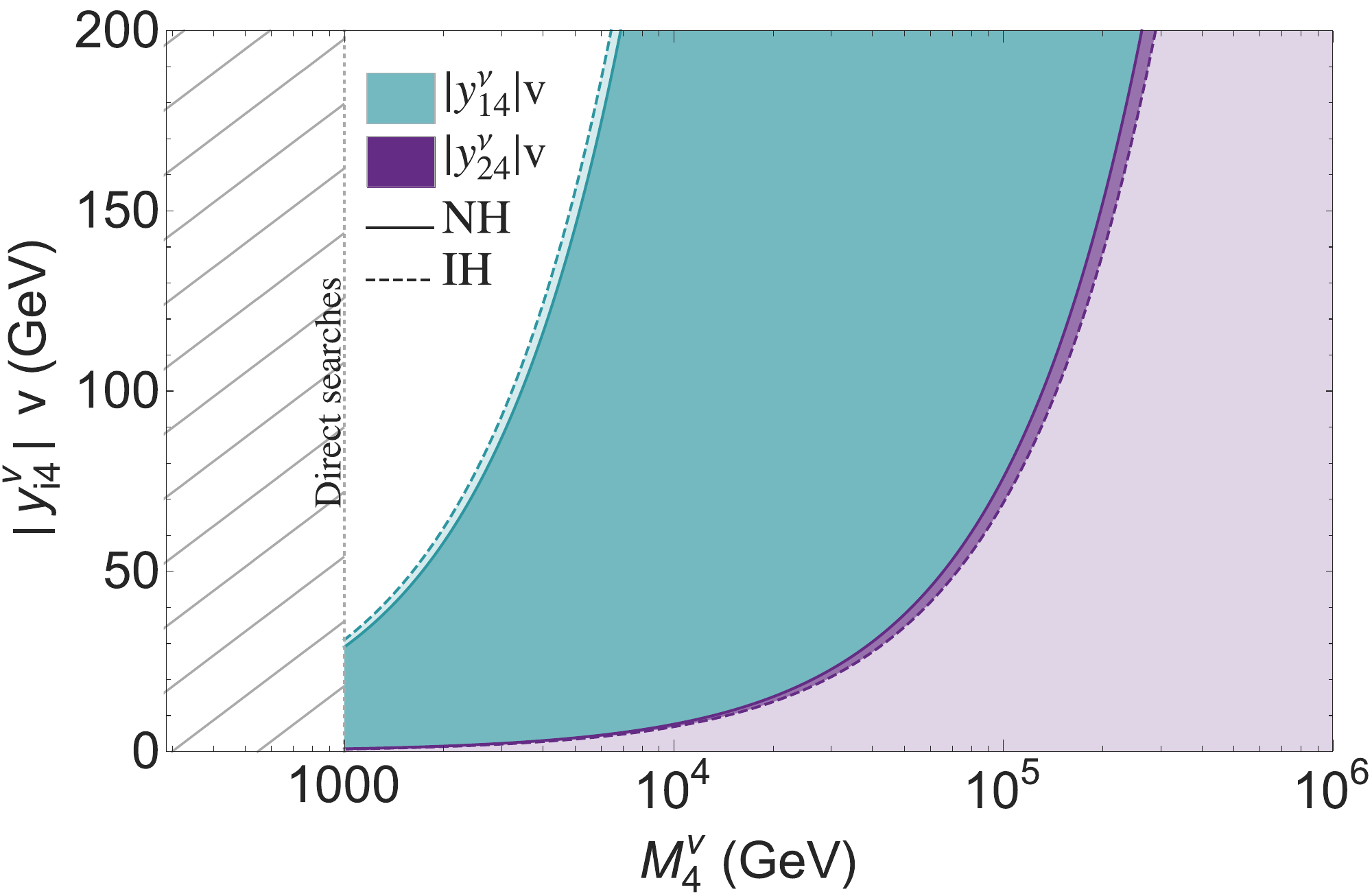}
\caption{Allowed region of the free parameters of the general scenario when the present bounds~\cite{Fernandez-Martinez:2016lgt} on the non-unitarity of the leptonic mixing matrix are considered. The green (purple) area corresponds to $\vert y_{14}^\nu\vert$ ($\vert y_{24}^\nu\vert$), while the solid (dashed) line is the boundary for NH (IH). The hatched gray area is already excluded by direct searches~\cite{Aad:2015kqa}.}
\label{fig:Yuk_limit_gen}
\end{figure}

In Figure~\ref{fig:Yuk_limit_gen} the constraints that the present bounds on the non-unitarity of the PMNS matrix of Eq.~(\ref{eq:eta_bounds}) set on the free parameters of the general scenario are shown. The allowed region for the $\vert y_{14}^\nu\vert$ ($\vert y_{24}^\nu\vert$) as a function of the vector-like neutrino mass scale $M_4^\nu$ is shown in green (purple). The solid (dashed) line corresponds to the boundary of the allowed region for a NH (IH) in the light neutrino sector. The hatched gray area has been already excluded by direct searches at LHC~\cite{Aad:2015kqa}.\\

Finally, if an arbitrary number $n$ (with $n\geq 2$) of $N^c$ fields is introduced in the model, the $\eta$ matrix would be a completely generic Hermitian matrix described by 9 free parameters. These parameters are enough to reproduce the correct masses and mixings of the light neutrinos, and thus $\hat m$ and $\eta$ would be unrelated. That is, there would not be correlations among the Yukawa couplings, and no extra information on this vector-like model would be derived. Therefore, these scenarios are not further discussed in this work.

%%%%%%%%%%%%%%%%%%%%%%%
\section{Discussion and Conclusions}
\label{s:conclusions}
%%%%%%%%%%%%%%%%%%%%%%%

In this paper we have considered a new Weinberg operator for neutrino mass of the form $H_u\tilde{H_d}L_iL_j$ involving two 
different Higgs doublets $H_u, H_d$ with opposite hypercharge, where $\tilde{H_d}$ is the charge conjugated doublet.
We have considered a minimal model involving 
two Higgs doublets, charged under a $U(1)'$ gauge group which forbids the usual Weinberg operator but allows
the mixed one. The new Weinberg operator is then generated via 
two right-handed neutrinos oppositely charged under the $U(1)'$.
Such a version of the type I seesaw model, which we refer to as type Ib to distinguish it from the usual type Ia seesaw
mechanism which yields the usual Weinberg operator, allows the possibility of having potentially large violations
of unitarity of the leptonic mixing matrix whose bounds we have explored.  
However the minimal model only allows non-renormalisable Yukawa couplings for the charged fermions.

In the minimal model, the SM particle content is extended by two right-handed neutrinos $\nu^c$ and $\overline{\nu^c}$. 
%which might correspond to the components of a vector-like fourth family in the complete model. 
These heavy right-handed neutrinos are oppositely charged under the gauge $U(1)^\prime$. Since the SM has been extended with just two extra singlets, just two of the three light neutrinos will be massive. In order to reproduce the observed pattern of neutrino masses and mixings, a particular structure in the Yukawa couplings of the right-handed neutrinos is obtained. In particular, the Yukawa couplings of the heavy neutrino $\nu^c$ with the light neutrinos will be reconstructed from the elements of the PMNS mixing matrix, the neutrino mass splittings, and an overall scaling factor $y$. The presence of the heavy neutrinos generate deviations of unitarity in the leptonic mixing matrix, and thus, there would be an enhancement in the LFV processes due to the loss of the GIM cancellation. The stringent experimental limit on the LFV radiative decay $\mu\to e \gamma$ has been used to analyse the allowed parameter space of the free parameters $y$ and $M^\nu$ of the minimal scenario (see Figure~\ref{fig:Yuk_limit}).

%We have also considered a more general model which allows Yukawa couplings to be generated via a fourth vector-like family, charged under the $U(1)'$. We also considered the relaxation of the unitarity bound due to the further addition of a single right-handed neutrino, neutral under  $U(1)'$, yielding a usual type Ia seesaw contribution, in addition to the type Ib contribution.In the general scenario, the SM particle content is extended with a full vector-like fourth family (containing the $\nu^c$ and $\overline{\nu^c}$ of the minimal scenario) and one extra CP conjugated  right-handed neutrino $N^c$. 

We have also considered a more general model which allows all Yukawa couplings to be generated via a fourth vector-like family 
charged under the $U(1)'$ (including $\nu_4^c$ and $\overline{\nu_4^c}$ identified as $\nu^c$ and $\overline{\nu^c}$ of the minimal scenario). In addition, we considered the relaxation of the unitarity bound due to the further addition of one (or more) extra CP conjugated right-handed neutrino(s) $N^c$, neutral under  $U(1)'$, yielding a usual type Ia seesaw contribution, in addition to the type Ib contribution.
In this way, all the SM fermions acquire Dirac masses via effective Yukawa couplings with the fourth family. 

In the case of one additional $N^c$ (plus $\nu_4^c$ and $\overline{\nu_4^c}$),
the three heavy neutrinos generate masses for the three light neutrinos, and as a result, the strong correlations on the Yukawa couplings of the minimal scenario are relaxed. In particular, two of the three Yukawa couplings that enter in the description of the dim-6 effective operator ($\eta$) will be free, and the third one will be given by the other two, and the pattern of masses and mixings of the light sector. 
%The presence of the heavy neutrinos would induce deviations of unitarity in the leptonic mixing matrix of the CC interactions. These non-unitarity parametrised by $\eta$ would modify Electroweak and flavour precision observables. 
The non-unitarity of the leptonic mixing matrix, generated by the presence of the heavy neutrinos and parametrised by $\eta$, would modify Electroweak and flavour precision observables. And thus, the present bounds on the non-unitarity of the mixing matrix can be used to constrain regions of the parameter space of the two free Yukawa couplings and $M_4^\nu$ (see Figure~\ref{fig:Yuk_limit_gen}).

%Finally, we have seen that if $n\geq 2$ extra $N^c$ fields are introduced in the model, the correlations among the Yukawa couplings would be lost, and the coefficients of the dim-5 and dim-6 effective operators would be unrelated. Therefore, the deviations of unitarity would be described by a generic Hermitian matrix, and no extra information on the free parameters of the model would be obtained.

In conclusion we have considered a new Weinberg operator for neutrino mass and proposed a type Ib seesaw mechanism to account for it. While the minimal model is quite compact and constrained by unitarity, it is not complete since the charged fermion Yukawa couplings are non-renormalisable. In order to obtain a renormalisable explanation of such Yukawa couplings, we were led to introduce a fourth vector-like family, to which the singlet neutrinos of the minimal model belong, leading to possible connections with $R_{K^{(*)}}$ as well as collider implications
for the LHC.\\

%%%%%%%%%%%%%%%%%%%%%%%
\section*{Acknowledgements}
%%%%%%%%%%%%%%%%%%%%%%%
This project has received funding/support from the European Union's Horizon 2020 research and innovation programme under the Marie Sk\l odowska-Curie grant agreement No.\ 674896 and InvisiblesPlus RISE No.\ 690575. JHG warmly thanks Southampton University for its hospitality hosting him during the discussion and the completion of this work. SFK acknowledges the STFC Consolidated Grant ST/L000296/1.\\

\providecommand{\href}[2]{#2}\begingroup\raggedright\endgroup

\end{document}